# Deep Learning–Based Dose Prediction for Automated, Individualized Quality Assurance of Head and Neck Radiation Therapy Plans

Mary P. Gronberg[1,2], Beth M. Beadle[3], Adam S. Garden[4], Heath Skinner[5], Skylar Gay[1,2], Tucker Netherton[1,2], Wenhua Cao[1], Carlos E. Cardenas[6], Christine Chung[1], David T. Fuentes[2,7], Clifton D. Fuller[2,4], Rebecca M. Howell[1,2], Anuja Jhingran[4], Tze Yee Lim[1,2], Barbara Marquez[1,2], Raymond Mumme[1], Adenike M. Olanrewaju[1], Christine B. Peterson[2,8], Ivan Vazquez[1], Thomas J. Whitaker[1,2], Zachary Wooten[8,9], Ming Yang[1,2], Laurence E. Court[1,2]

[1] Department of Radiation Physics, The University of Texas MD Anderson Cancer Center, Houston, Texas, USA

[2] The University of Texas MD Anderson Cancer Center UTHealth Houston Graduate School of Biomedical Sciences, Houston, Texas, USA

[3] Department of Radiation Oncology, Stanford University, Stanford, California, USA

[4] Department of Radiation Oncology, The University of Texas MD Anderson Cancer Center, Houston, Texas, USA

[5] Department of Radiation Oncology, University of Pittsburgh, Pittsburgh, Pennsylvania, USA

[6] Department of Radiation Oncology, The University of Alabama at Birmingham, Birmingham, Alabama, USA

[7] Department of Imaging Physics, The University of Texas MD Anderson Cancer Center, Houston, Texas, USA

[8] Department of Biostatistics, The University of Texas MD Anderson Cancer Center, Houston, Texas, USA

[9] Department of Statistics, Rice University, Houston, Texas, USA

Corresponding Author: Mary P. Gronberg
Email: mpeters1@mdanderson.org

**ABSTRACT**

**Purpose:** This study aimed to use deep learning–based dose prediction to assess head and neck (HN) plan quality and identify suboptimal plans.

**Methods:** A total of 245 volumetric modulated arc therapy HN plans were created using RapidPlan knowledge-based planning (KBP). A subset of 112 high-quality plans was selected under the supervision of an HN radiation oncologist. We trained a 3D Dense Dilated U-Net architecture to predict 3-dimensional dose distributions using 3-fold cross-validation on 90 plans. Model inputs included computed tomography images, target prescriptions, and contours for targets and organs at risk (OARs). The model's performance was assessed on the remaining 22 test plans. We then tested the application of the dose prediction model for automated review of plan quality. Dose distributions were predicted on 14 clinical plans. The predicted versus clinical OAR dose metrics were compared to flag OARs with suboptimal normal tissue sparing using a 2 Gy dose difference or 3% dose-volume threshold. OAR flags were compared with manual flags by 3 HN radiation oncologists.

**Results:** The predicted dose distributions were of comparable quality to the KBP plans. The differences between the predicted and KBP-planned $D_{1\%}$, $D_{95\%}$, and $D_{99\%}$ across the targets were within −2.53%±1.34%, −0.42%±1.27%, and −0.12%±1.97%, respectively, and the OAR mean and maximum doses were within −0.33±1.40 Gy and −0.96±2.08 Gy, respectively. For the plan quality assessment study, radiation oncologists flagged 47 OARs for possible plan improvement. There was high interphysician variability; 83% of physician-flagged OARs were flagged by only one of 3 physicians. The comparative dose prediction model flagged 63 OARs, including 30 of 47 physician-flagged OARs.

**Conclusion:** Deep learning can predict high-quality dose distributions, which can be used as comparative dose distributions for automated, individualized assessment of HN plan quality.

## 1. INTRODUCTION

Radiation therapy is a cornerstone for the treatment of patients with head and neck (HN) cancers, with almost 75% of patients benefiting from its use[1,2]. HN radiation plans are complex and challenging to create compared with many primary cancers, as they often have multiple targets and dose levels. Moreover, critical normal structures are difficult to spare from radiation dose as they are near the targets. The challenging nature of HN treatment planning is often further complicated by a lack of institutional experience. Owing to the low incidence of HN cancers in the United States, which make up approximately 3% of all new cancer cases[3], many radiation oncologists have less consistent experience in treating patients with HN cancer, which negatively affects patient survival[4-6].

HN plan quality is often assessed using scorecards (ie, institution-specific dose-volume objectives and constraints displayed in the form of a checklist). The normal tissue constraints are typically well-established literature-based and protocol-based thresholds for severe toxicity[7,8]. Meeting planning goals indicates a certain level of plan safety, but not optimality. Many studies have shown a correlation between radiation doses to critical organs at risk (OARs) and side-effect severity[9-13], supporting the benefit of continued treatment planning iterations to further spare normal tissue OARs.

During the past several years, researchers have used deep learning to predict 3-dimensional (3D) patient-specific dose distributions for HN plans[14-20]. These deep learning algorithms receive similar inputs as human planners, including the patient's computed tomography (CT) images, normal tissue contours, target delineations, and dose levels. By using high-quality plans during training, the models can learn to accurately predict dose distributions with high degrees of normal tissue sparing for new patients. Predicted doses can be used to guide plan optimization as part of an automated planning process[21,22].

In this study, we investigated a different use of deep learning–based dose predictions as a plan quality assessment (QA) tool. The development of automated plan QA tools can help reduce disparities in access to high-quality radiation therapy. Most patients are treated at community clinics, where radiation oncologists treat and dosimetry and physics staff plan all disease sites. In contrast, radiation oncology physician, dosimetry, and physics teams at larger centers often have disease-site specializations and routinely implement site-specific expert review programs, where each treatment plan is reviewed by other radiation oncologists with the same disease specialization. Providing access to plan QA tools that were trained with high-quality plans that passed specialized-peer review can provide automated, expert peer review, without needing actual access to teams of specialized radiation oncologists.

This study aimed (1) to predict optimum achievable radiation therapy plans for patients with HN cancer using deep learning and (2) to evaluate the usability of predicted dose distributions as comparative dose distributions to assess the quality of HN radiation therapy. To our knowledge, this is the first study to use deep learning–based 3D dose prediction for plan QA. We present the results of our plan QA tool and compare its performance with HN radiation oncologist peer review.

## 2. METHODS AND MATERIALS

### 2.1 Deep learning–based dose prediction for head and neck radiation therapy

In this study, 245 volumetric modulated arc therapy HN plans were created using a previously validated automated planning approach[23] based on RapidPlan (Varian Medical Systems, Palo Alto, CA) knowledge-based planning (KBP). The collection and use of this data were performed under a protocol approved by the University of Texas MD Anderson Cancer Center institutional review board. The KBP plans were generated using clinical target volumes and autocontoured OARs, generated with an autocontouring tool based on convolutional neural networks[24]. This semiautomated planning and contouring approach was selected to ensure consistency of the dose prediction model training data. A subset of 112 high-quality plans was selected based on institutional preferences and achievable normal tissue sparing under the supervision of an HN radiation oncologist to train and test the deep learning dose prediction model. The data set included both definitive and postoperative radiation therapy cases with varied dose prescriptions to multiple targets. Targeted sites included: oral cavity, oropharynx, larynx, salivary gland, hypopharynx, thyroid, paranasal sinuses, and the neck. The cases also included heterogeneity with regards to the neck as in some cases only one side of the neck was treated, while in others both sides of the neck were targeted.

We used a 3D Dense Dilated U-Net architecture with a weighted mean squared error loss function modeled off inverse treatment planning, which assigns a higher weight to the dose distribution within the targets and serial OARs. This architecture was our top-performing architecture[19] from the American Association of Physicists in Medicine Open Knowledge-Based

Planning (OpenKBP) Grand Challenge[25]. The model receives 24 inputs for each patient including the normalized CT simulation scan (voxel intensities were cropped to the range [-1000,1000] and rescaled to the range [0,1]), the target contours and their prescriptions (target array of prescription values), and the OAR contours (image masks) including the body, brain, brain stem, oral cavity, optic chiasm, left and right cochlea, esophagus, left and right eye, left and right submandibular gland, larynx, mandible, left and right lens, left and right optic nerve, left and right parotid, spinal cord, and vertebral column.

The data set was split (4:1) into 90 plans for training and 22 plans for testing. The 90 training plans were split into 3 equal folds of 30 plans. Three-fold cross-validation was used for training, meaning that each of 3 models was trained with 2-folds of data and validated on the remaining fold. Because the head and neck region is symmetrical about the sagittal plane, we were able to augment the training data by flipping the patient images and dose distributions left-right for each plan in the training data set. Training with both the original and flipped data enabled us to effectively double our training data set. Each model was trained using image "patches", cropped portions of the contoured CT of size (64, 64, 64) and the corresponding cropped dose distribution. An NVIDIA Tesla V100 graphics processing units with 16 GB memory were used to train the models using a batch size of 8 with a maximum of 1000 epochs using early stopping. The Adam optimization algorithm was used with an initial learning rate of 0.001, set to decrease by half each time the loss did not improve over 55 epochs.

The model performance was assessed on the test set of 22 plans. Test set predictions were an average of the dose predictions from the 3 final models, one for each fold of the training and validation data. For each plan in the test set, the images were split into overlapping patches of size (64, 64, 64). Patches were generated using a stride of 16 along each axis. For a given plan and model, every patch was input into the model, and the predicted dose within the body contour was averaged across the patches. We performed a paired two one-sided *t* test to assess the equivalence of the predicted and KBP-planned mean doses under an equivalence bound of 1 Gy. $\Delta D_{V\%}(\%)$, the percentage error in predicted target metrics at $V\%$ level, were calculated using the following equation:

$$\Delta D_{V\%}(\%) = \frac{D_{V\%}(predicted) - D_{V\%}(KBP)}{D_{V\%}(KBP)} * 100\%, \tag{1}$$

where $D_{V\%}(predicted)$ and $D_{V\%}(KBP)$ are the least dose that the hottest $V\%$ of the target volume receives for the predicted and KBP-planned dose distributions, respectively. $\Delta D_{mean}$, the error in predicted OAR mean doses, and $\Delta D_{max}$, the error in predicted OAR maximum doses, were calculated using Equations 2 and 3. $D_{mean}(predicted)$ and $D_{max}(predicted)$ are the mean and maximum doses, respectively, of the predicted dose distributions within an OAR. Similarly, $D_{mean}(KBP)$ and $D_{max}(KBP)$ are the mean and maximum doses, respectively, of the KBP-planned dose distributions within an OAR.

$$\Delta D_{mean} = D_{mean}(predicted) - D_{mean}(KBP) \tag{2}$$

$$\Delta D_{max} = D_{max}(predicted) - D_{max}(KBP) \tag{3}$$

## 2.2 Automated plan quality assessment

Being cognizant of physician time, a subset of the 22 cases in the dose prediction test set was selected for the plan quality assessment study. Fourteen cases were selected as being most representative of the training data set. For this plan quality assessment study, the manually created clinical plans were assessed, not the KBP plans. The data set consisted of oral cavity, oropharynx, and larynx plans with equal numbers of postoperative and definitive cases. Three HN radiation oncologists from different institutions were asked to independently review each clinical plan, flagging OARs that would benefit from further sparing. For each of the 14 clinical plans, the deep learning dose prediction model ensemble was used to predict the 3D dose distribution. The OAR contours that were input into the dose prediction model and used for the plan quality assessment were the clinical contours, with autocontours used for OARs excluded from the clinical contours. The predicted and clinical dose distributions were compared within each OAR using the relevant OAR dose-volume histogram (DVH) metrics: $D_{mean}$ for the esophagus, larynx, oral cavity, parotids, and submandibular glands; $D_{max}$ for the brain, brain stem, chiasm, cochleas, eyes, lenses, mandible, optic nerves, and spinal cord; $V_{45Gy}$ and $V_{54Gy}$ for the esophagus; $V_{40Gy}$, $V_{60Gy}$, and $V_{70Gy}$ for the mandible; and $V_{30Gy}$ for the parotids. The differences in predicted and clinical OAR mean doses, $\Delta D_{mean}$, and maximum doses, $\Delta D_{max}$, were calculated using Equations 4 and 5, where $D_{mean}(clinical)$ and $D_{max}(clinical)$ are the mean and maximum doses, respectively, of the clinical dose distributions within an OAR.

$$\Delta D_{mean} = D_{mean}(predicted) - D_{mean}(clinical) \qquad (4)$$

$$\Delta D_{max} = D_{max}(predicted) - D_{max}(clinical) \qquad (5)$$

The differences in OAR DVH metrics at $D_{Gy}$ dose level, $\Delta V_{DGy}$, were calculated using

$$\Delta V_{DGy} = V_{DGy}(predicted) - V_{DGy}(clinical) \qquad (6)$$

where $V_{DGy}(predicted)$ and $V_{DGy}(clinical)$ are the relative volume of an OAR receiving a dose of $D_{Gy}$ or more for the predicted and clinical dose distributions, respectively. OARs were flagged for improvement using a threshold of a 2 Gy dose difference or 3% dose-volume difference for the OAR dose-volume metrics. The OARs flagged by the dose prediction algorithm were compared with those flagged by the physicians.

## 3. RESULTS

### 3.1 Deep learning–based dose prediction for head and neck radiation therapy

The predicted dose distributions closely resembled the KBP-planned dose distributions for both the postoperative and definitive cases in the test set. Figure 1 compares predicted and KBP-planned dose distributions and DVHs for an example postoperative case in the test set. The predicted dose distribution had good conformity and homogeneity for all 3 target levels. The predicted normal tissue sparing and dose falloff from the targets were comparable to those of the knowledge-based plan, indicated by the closely matching isodose lines. Similarly, there was excellent agreement between the predicted (dashed) and KBP-planned (solid) DVH lines, as shown in Figure 1B. Although the postoperative cases had consistent dose levels of 60, 57, and 54 Gy, the dose levels for the definitive cases varied. Despite this added complexity, our model achieved accurate predictions for definitive cases. Figure 2 compares the predicted and KBP-planned dose distributions for an example definitive case in the test set, showing good agreement in target coverage and normal tissue sparing.

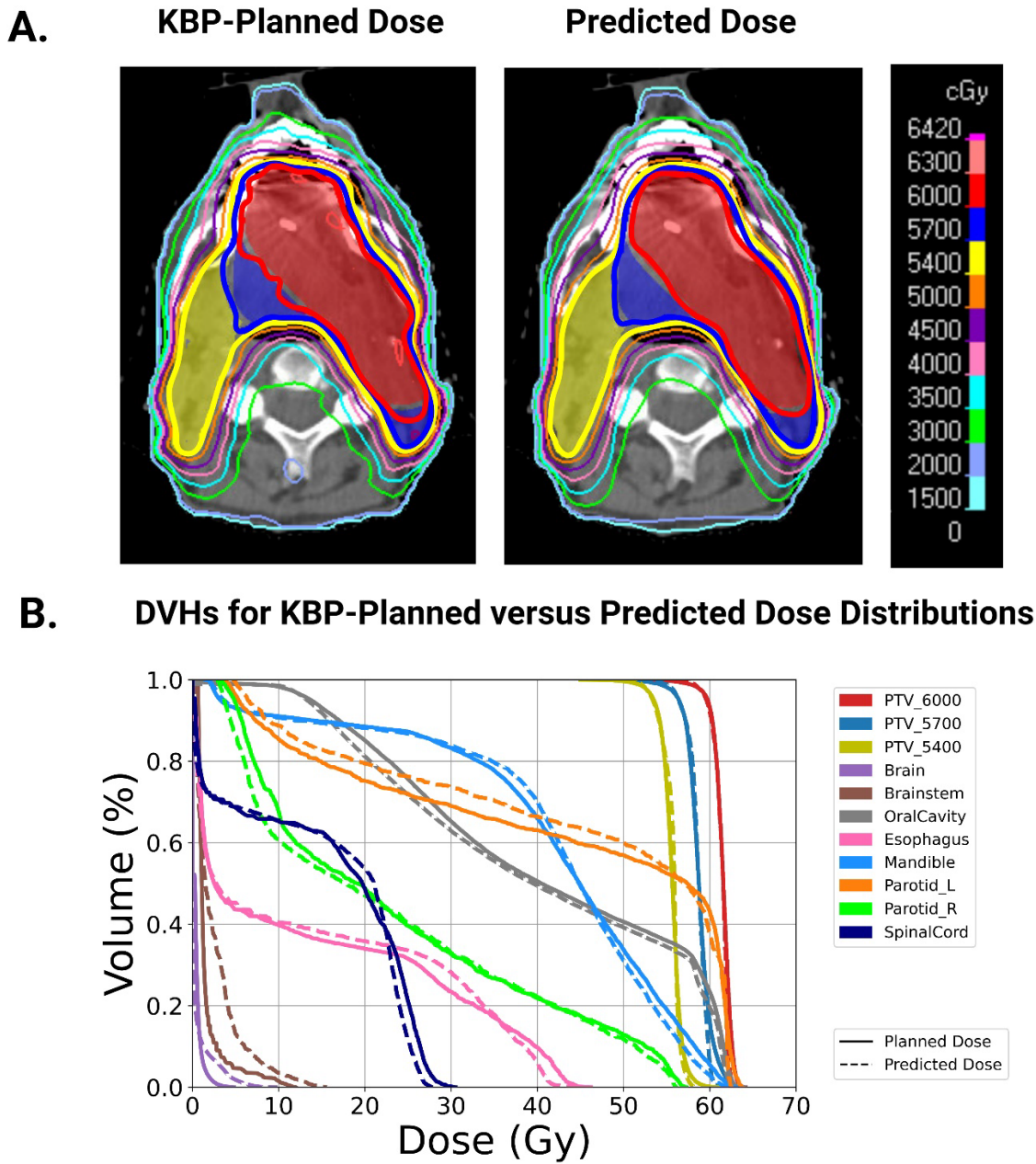

**Figure 1.** (A) Comparison of the knowledge-based planning (KBP)-planned (left) and predicted (right) dose distributions for a postoperative case in the test set. Shading in red indicates high-dose planning target volume (PTV), blue indicates intermediate-dose PTV, and yellow indicates low-dose PTV. (B) Comparison of KBP-planned (solid) and predicted (dashed) dose-volume histograms (DVHs) for the same case.

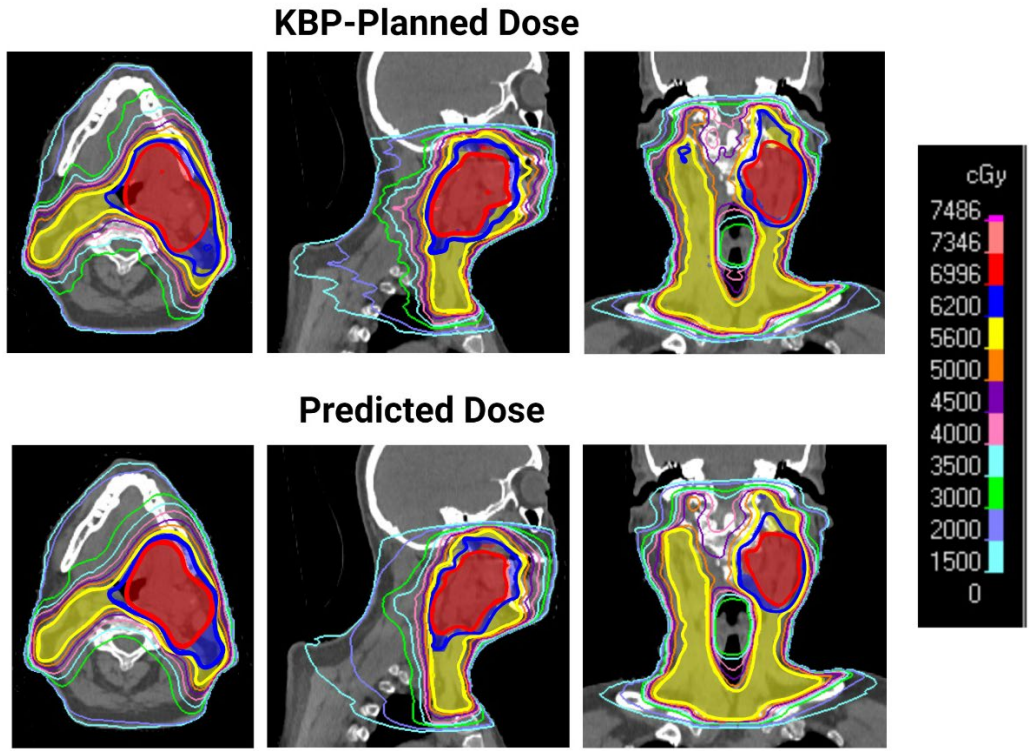

**Figure 2.** Comparison of the knowledge-based planning (KBP)-planned (top row) and predicted (bottom row) dose distributions for a definitive case in the test set.

The results from our paired two one-sided *t* test showed that the model predicted and KBP-planned dose distributions of the test set were equivalent (*P* <0.0001). The mean voxel-wise dose difference was −0.31±0.36(SD) Gy, with the predicted doses being slightly lower than the KBP doses. Figure 3 displays the percentage error in predicted target metrics. The predicted target DVH metrics of $D_{1\%}$, $D_{95\%}$, and $D_{99\%}$ averaged over all target levels were within −2.53%±1.34%, −0.42%±1.27%, and −0.12%±1.97% of the KBP target metrics, respectively.

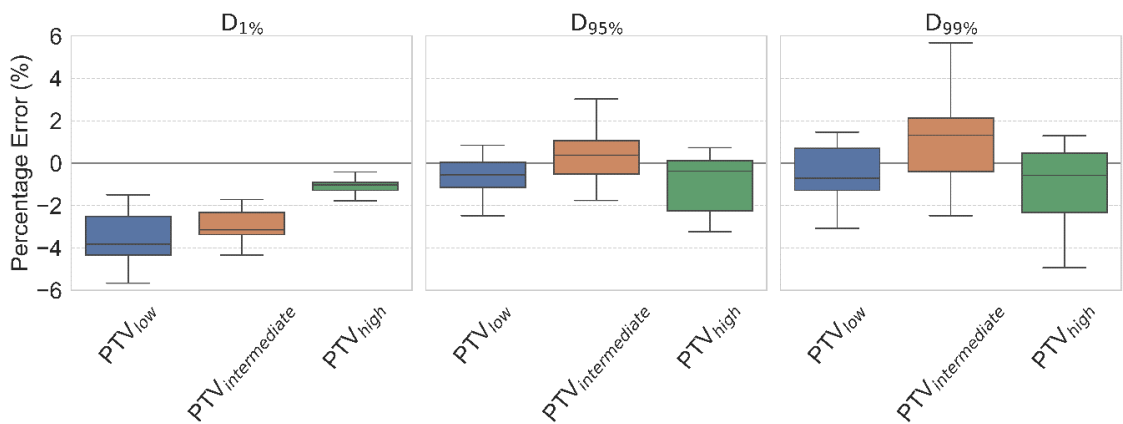

**Figure 3.** The percent error between the knowledge-based planned and predicted target metrics of $D_{1\%}$, $D_{95\%}$, and $D_{99\%}$ for each planning target volume (PTV) in the test data set.

Figure 4 displays the error in predicted OAR metrics. Predicted OAR mean doses were within −0.33±1.40 Gy and predicted OAR maximum doses were within −0.96±2.08 Gy of the KBP-planned OAR doses.

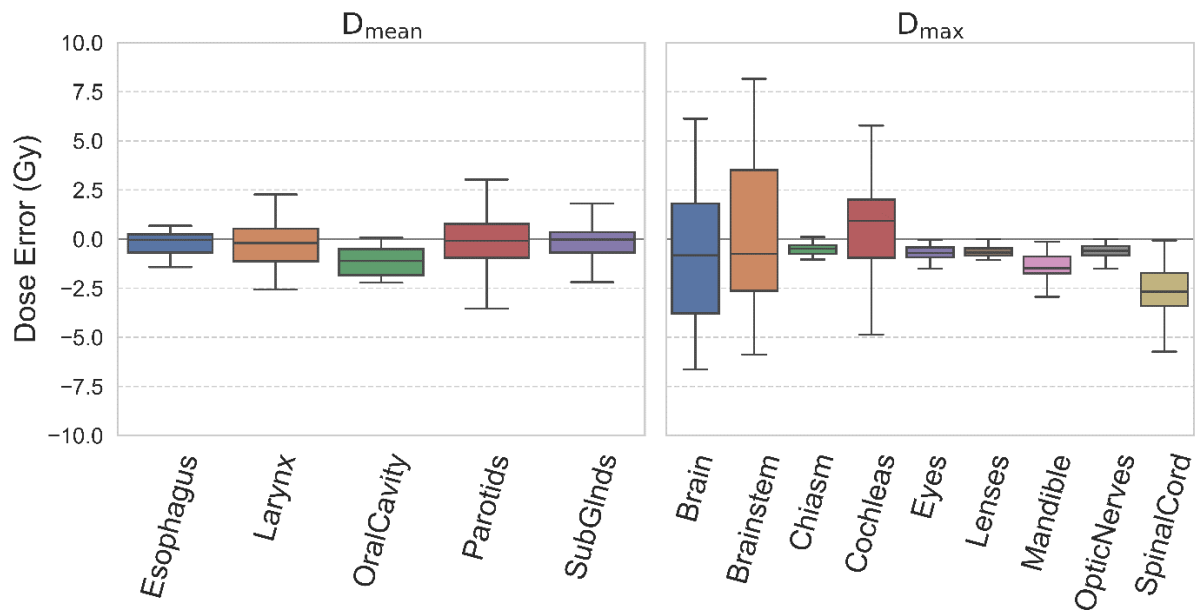

**Figure 4.** The dose difference between the knowledge-based planned and predicted $D_{mean}$ and $D_{max}$ for organs at risk in the test data set. *Abbreviation:* SubGlnds = submandibular glands.

### 3.2 Automated plan quality assessment

The OARs flagged by physicians were compared to those flagged by the comparative dose prediction model. Forty-seven OARs were flagged by at least one physician, but 83% of the physician-flagged OARs were flagged by only 1 of the 3 physicians. Each HN physician in our study had different OAR flagging tendencies. Physician A had the most physician flags (38/47 flags) with a relatively even distribution of flags among all OARs. Physician B had 9 total flags, with 7 of 9 flags on the mandible and parotids, and physician C had 6 flags, with 4 of 6 flags on the esophagus. The comparative dose prediction model flagged 63 OARs, as it predicted that the examined OAR dose-volume metrics (Table 1, column 2) could be reduced by at least 2 Gy or 3%. The comparative dose prediction model flagged 30 of 47 physician-flagged OARs. Table 1 details the number of plans flagged for suboptimal sparing per OAR by the physicians, the model, and both the physician(s) and the model.

**Table 1.** Comparison of physician flags and dose prediction–based flags of suboptimal OAR sparing for 14 clinical plans.

| OAR | OAR Metrics Compared for Dose Prediction–Based Auto-Flags | Number of Plans Flagged by Physician(s) | Number of Plans Flagged by Comparative Dose Prediction Model | Number of Plans Flagged by Both Physician(s) and Comparative Dose Prediction Model |
|---|---|---|---|---|
| Brain | $D_{max}$ | 6 | 10 | 4 |
| Brain stem | $D_{max}$ | 7 | 4 | 2 |
| Chiasm | $D_{max}$ | 0 | 0 | 0 |
| Cochleas | $D_{max}$ | 1 | 4 | 1 |
| Esophagus | $D_{mean}, V_{45Gy}, V_{54Gy}$ | 7 | 4 | 2 |
| Eyes | $D_{max}$ | 0 | 0 | 0 |
| Larynx | $D_{mean}$ | 2 | 5 | 1 |
| Lenses | $D_{max}$ | 0 | 0 | 0 |
| Mandible | $D_{max}, V_{40Gy}, V_{60Gy}, V_{70Gy}$ | 6 | 13 | 6 |
| Optic Nerves | $D_{max}$ | 0 | 0 | 0 |
| Oral Cavity | $D_{mean}$ | 4 | 6 | 4 |
| Parotids | $D_{mean}, V_{30Gy}$ | 5 | 4 | 2 |
| Spinal Cord | $D_{max}$ | 8 | 8 | 7 |
| Submandibular Glands | $D_{mean}$ | 1 | 5 | 1 |
| *Abbreviation:* OAR = organ at risk. | | | | |

The results for the oral cavity are presented to provide a deeper understanding of our methodology and findings. A scatterplot comparing model-predicted versus clinical mean dose to the oral cavity is displayed in Figure 5. The deep learning-based dose prediction model predicted that further sparing of the oral cavity was achievable for 6 of the 14 cases. All plans with suboptimal oral cavity sparing that were flagged by physicians were also flagged by the dose prediction model. Moreover, the dose prediction model identified 2 additional plans that were not flagged by physicians that could benefit from further sparing of the oral cavity.

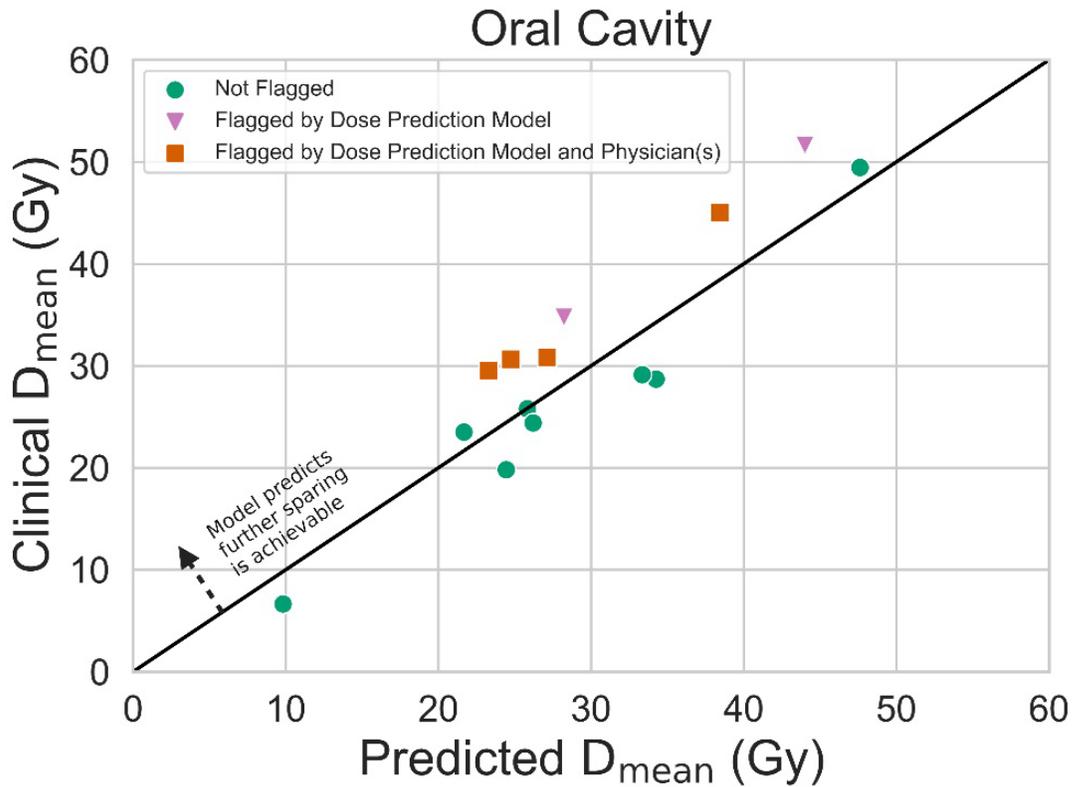

**Figure 5.** Scatterplot comparing model predicted versus clinical mean dose to the oral cavity. Data points *above* the solid black line indicate cases for which the deep learning–based dose prediction model predicted that further sparing of the oral cavity was achievable. Four of these 6 cases were likewise flagged by at least 1 physician.

Examples of agreement and disagreement in plan QA between the physicians and the comparative dose prediction model are shown in Figure 6. In one case, the spinal cord was flagged by both a physician and the comparative dose prediction model (Figure 6A). In the predicted dose distribution, the spinal cord is spared by the 30 Gy isodose line. For this patient, the deep learning model predicted a spinal cord maximum dose of 28.66 Gy, a reduction by more than 12 Gy of the clinical spinal cord maximum dose of 41.41 Gy. In another case, the oral cavity was flagged by the dose prediction model but not by the physicians (Figure 6B). In the clinical plan, there is a very sharp dose gradient falloff from the target to the brain stem. The predicted dose distribution suggests that by relaxing the constraint to the brain stem, the oral cavity outside of the target could be better spared by the 30 Gy isodose line. Additionally, the dose prediction showed that relaxing the constraint to the brain stem could also reduce the dose spillage to the back of the neck, which was noted to be of concern by 2 physicians. In a third case, the esophagus was flagged for further sparing by a physician but not flagged by the dose prediction model (Figure 6C). The physician indicated that the esophagus could be spared by the 40 Gy isodose line. In both the clinical and predicted dose distributions, the 40 Gy isodose line was being pushed off the airway but not the esophagus.

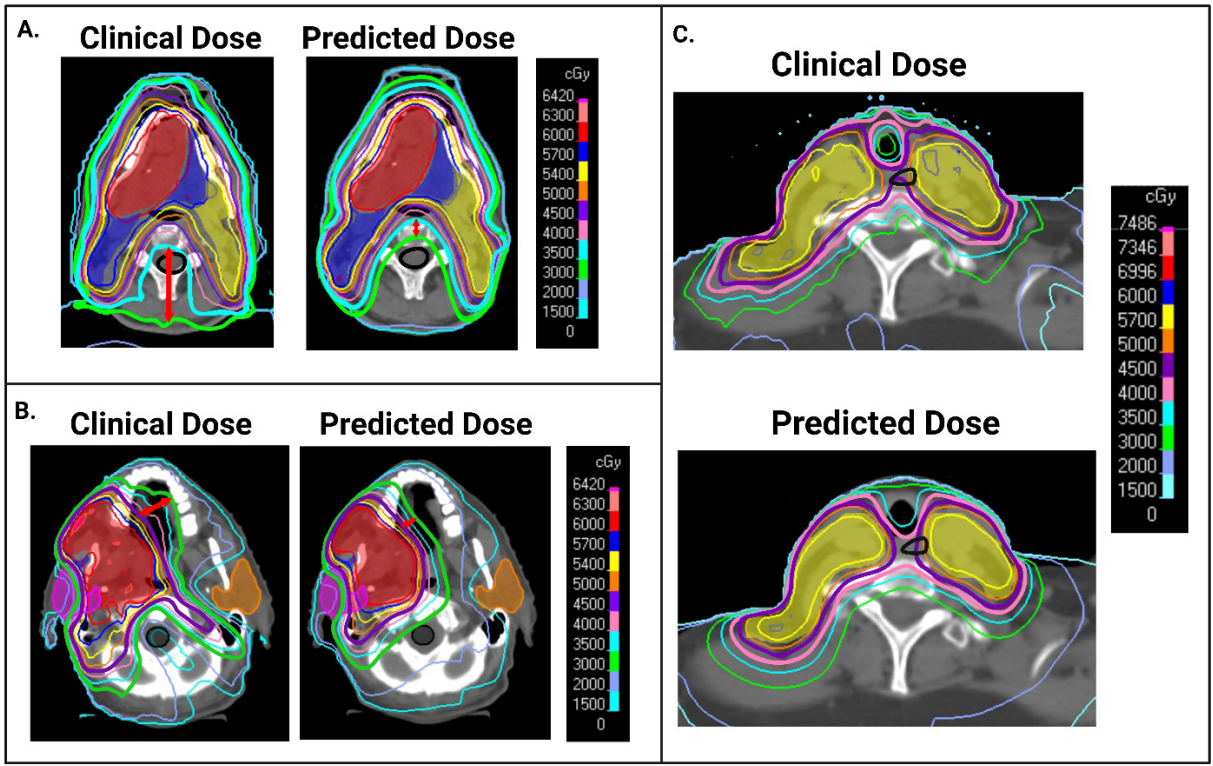

**Figure 6.** Example cases displaying agreement and disagreement in organ at risk flags for further sparing between head and neck radiation oncologists and the comparative dose prediction model. (A) The spinal cord (outlined in black) was flagged by both the dose prediction model and a physician as possible for further sparing. The dose prediction model predicted that the spinal cord could be spared by the 30 Gy isodose line (lime green). (B) The oral cavity was flagged by the dose prediction model but not by a physician. The predicted dose distribution indicated that the oral cavity could be further spared by the 30 Gy isodose line (lime green) by relaxing the constraint to the brain stem (outlined in black). (C) The esophagus (outlined in black) was flagged by a physician but not by the dose prediction model as needing further sparing. The physician indicated that the esophagus could be spared by the 40 Gy isodose line (pink).

For most OARs, dose prediction–based flagging accurately captured physician-flagged OARs. Excluding the esophagus and brain stem, 79% of the physician-flagged OARs were flagged by the comparative dose prediction model. Figure 7 displays the dose difference in predicted versus clinical OAR metrics for each OAR. The top 3 OARs flagged by the comparative dose prediction model were the mandible, brain, and spinal cord, with 13, 10, and 8 plans flagged, respectively. On average, the dose prediction model predicted a possible reduction in maximum dose to the mandible, brain, and spinal cord of 2.10±3.83, 6.04±7.56, and 3.36±5.02 Gy, respectively. The top 3 OARs flagged by physicians were the spinal cord, esophagus, and brain stem, with 8, 7, and 7 plans flagged, respectively. Although there was good agreement between physician and dose prediction–based flagging for the spinal cord, dose prediction was less accurate at detecting opportunities for further sparing of the esophagus and brain stem. In fact, the dose prediction model typically predicted worse sparing of these structures. On average, the mean dose to the esophagus and maximum dose to the brain stem were 5.34±3.90 Gy and 2.44±8.34 Gy higher, respectively, in the predicted dose distribution than in the clinical plan.

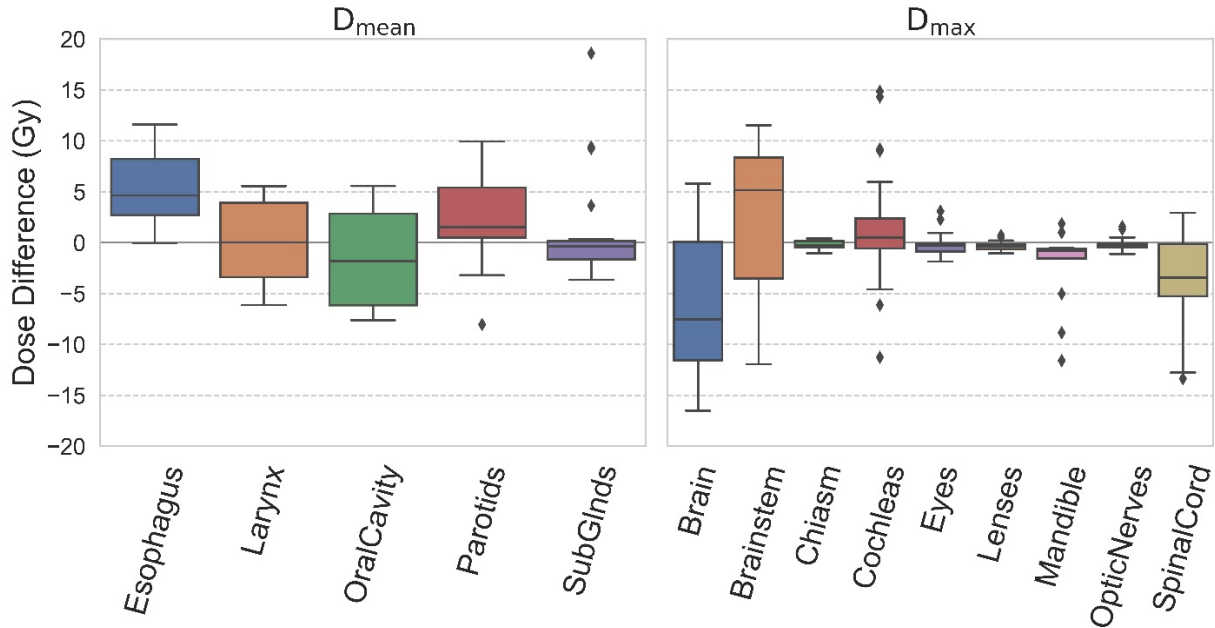

**Figure 7.** Difference in predicted and clinical $D_{mean}$ and $D_{max}$ for organs at risk across the 14 plans. Negative dose differences indicate that the deep learning dose prediction model predicted that further sparing was achievable. *Abbreviation:* SubGlnds = submandibular glands.

## 4. DISCUSSION

In this study, we developed an automated plan QA tool for HN cancers using a deep learning approach and benchmarked its performance against 3 HN physicians. To our knowledge, this is the first study to use deep learning–based 3D dose prediction for plan QA. This is an important extension to prior literature, that uses DVH prediction for plan QA[26,27]. 3D dose prediction has advantages compared to DVH prediction regarding plan QA. If an OAR is flagged as possible for further sparing, the predicted 3D dose distribution can show the reviewer how to modify the plan and what potential tradeoffs may result. In contrast, the DVH prediction can only show the reviewer the possible DVH curve for that OAR. Also, tradeoffs are not considered with DVH prediction, as DVH models are specific to an OAR, unlike 3D dose prediction models, which consider the entire patient's geometry.

An unexpected finding of this study was high interphysician variability in plan review, with 83% of physician-flagged OARs being flagged by only one of the 3 HN radiation oncologists. This finding is in line with those of Ventura et al[28], which compared the plan preferences of 3 radiation oncologists with their plan QA tool. Physicians were asked to select the preferred plan for 20 paired cases. All 3 physicians selected the same plan only half of the time. Interphysician variability in our study may be attributed to differences in physician preferences. The variability of flagging among the 3 physicians was quite striking both with regards to the number of flags (38 for one physician compared to 15 combined for the other 2 physicians), and structures flagged, with one physician having multiple OARs flagged while the other 2 were more focused on specific and differing structures. These differences in sparing preferences may reflect their different training backgrounds or institutional practices. They could also be the result of their experience with the symptom burden of and the prevalence of radiation side effects in their own patients. For example, Physician B may prioritize parotid and mandibular sparing to reduce the severity and prevalence of xerostomia and osteoradionecrosis among his/her patients. Similarly, Physician C may prioritize sparing the esophagus and other swallowing structures to reduce the risk of dysphagia. Furthermore, the flagging patterns may reflect what the physicians believe is feasible, based on their experience with prior plan quality, and not necessarily a preference; again, this would be based on their individual experience and the quality of the plans they have experienced. The comparative dose prediction model was able to capture the flagging patterns of all 3 physicians, correctly identifying 63%, 78%, and 67% of physician A, B, and C's flags, suggesting that the implementation of deep learning dose prediction tools in current plan review practices may help reduce observed interphysician variability and improve overall plan quality.

One limitation of using 3D dose prediction to automate plan review is that the quality of the predictions is dependent on the quality of the plans used to train the deep learning model. Despite great effort and physician involvement in the development of the KBP model that was used to generate the training data set for the deep learning model, as well as careful manual review and filtering of the training data set under physician supervision, the deep learning model still predicted suboptimal sparing of the esophagus for some of the clinical plans. To improve the performance of the model in identifying suboptimal sparing of these OARs, the deep learning algorithm could be retrained with plans that better spare the esophagus. Another limitation of this study was in the selection of an automated flagging threshold. A flagging threshold of a 2 Gy dose difference or 3% dose-volume difference for OAR dose-volume metrics was selected for flagging OARs with suboptimal sparing for automated QA using dose prediction. More sophisticated thresholds could incorporate normal tissue complication probabilities for each OAR, such that plans are flagged for review when the predicted dose distribution indicates that the patient is being exposed to a certain amount of excess risk of toxicity. In future work, we will investigate potential toxicity reductions from automated plan review with deep learning–based dose prediction using retrospective clinical trial data.

## 5. CONCLUSION

We have demonstrated that deep learning can accurately predict high-quality dose distributions for HN cancer patients, which can be used as comparative dose distributions for automated, individualized assessment of HN plan quality. This automated plan QA tool can be used to support plan review and improve the overall quality and consistency of radiation therapy, especially when clinicians do not have access to other subspecialized radiation oncologists. For clinics with adequate subspecialized radiation oncologists for meaningful plan review, automated plan review can be used to triage plans for review so that physicians can give more attention to plans that would benefit the most from review. Automated plan review also has the potential to assist clinics around the world with the safe and effective adoption of new technology (eg, transitioning from 3D conformal radiation therapy to intensity modulated radiation therapy) by providing data-driven decision support.

## 6. ACKNOWLEDGEMENTS

The authors acknowledge the support of the High Performance Computing for Research facility at The University of Texas MD Anderson Cancer Center for providing computational resources that have contributed to the research results reported in this paper. The manuscript was edited by Sarah Bronson of the Research Medical Library at MD Anderson.

## 7. SOURCES OF SUPPORT

Funding for this project was provided by the National Center for Advancing Translational Sciences of the National Institutes of Health under Award Numbers TL1TR003169 and UL1TR003167, the Cancer Prevention and Research Institute of Texas, and the American Legion Auxiliary Fellowship in Cancer Research. The content is solely the responsibility of the authors and does not necessarily represent the official views of the National Institutes of Health.

## 8. DISCLOSURES

Authors are involved with the Radiation Planning Assistant, which receives funding from the National Cancer Institute (NCI), Cancer Prevention and Research Institute of Texas (CPRIT), Wellcome Trust, and Varian Medical Systems. Mary Gronberg received funding and travel support from the National Center for Advancing Translational Sciences of the National Institutes of Health (NIH) under Award Numbers TL1TR003169 and UL1TR003167. The content is solely the responsibility of the authors and does not necessarily represent the official views of the NIH. Mary Gronberg was supported by the American Legion Auxiliary Fellowship in Cancer Research. Mary Gronberg received travel support for attending meetings from the University of Texas MD Anderson Cancer Center UTHealth Houston Graduate School of Biomedical Sciences Travel Award. Unrelated to this work, Mary Gronberg received support from the Linda M. Wells GSBS Outreach Award.  the course of this work, Mary Gronberg served on committees for the American Association of Physicists in Medicine (AAPM). Skylar Gay was supported by the 2021 AAPM Radiological Society of North America (RSNA) Graduate Fellowship. Dr. Netherton was partially supported by CPRIT. Dr. Fuller receives funding and salary support unrelated to this project during the period of study execution from: the NIH National Institute of Biomedical Imaging and Bioengineering (NIBIB) Research Education Programs for Residents and Clinical Fellows Grant (R25EB025787-01); the National Institute for Dental and Craniofacial Research (NIDCR) Establishing Outcome Measures Award (1R01DE025248/R56DE025248); the NIDCR Academic Industrial Partnership Grant (R01DE028290); the NIH NIDCR Exploratory/Developmental Research Grant Program (R21DE031082); the NCI Early Phase Clinical Trials in Imaging and Image-Guided Interventions Program  (1R01CA218148); the NCI Parent Research Project Grant (R01CA258827); an NIH/NCI Cancer Center Support Grant (CCSG) Pilot Research Program Award from the UT MD Anderson CCSG Radiation Oncology and Cancer Imaging Program (P30CA016672); an NIH/NCI Head and Neck Specialized Programs of Research Excellence (SPORE) Developmental Research Program Award (P50CA097007);  the NIH Big Data to Knowledge (BD2K) Program of the NCI Early Stage Development of Technologies in Biomedical Computing, Informatics, and Big Data Science Award (1R01CA214825); the NSF, Division of Mathematical Sciences, Joint NIH/NSF Initiative on Quantitative Approaches to Biomedical Big Data (QuBBD) Grant (NSF 1557679); the NSF Division of Civil, Mechanical, and Manufacturing Innovation (CMMI) grant (NSF 1933369); the NSF/NCI Smart Connected Health Program (R01CA257814); a Small Business Innovation Research Grant Program sub-award from Oncospace, Inc. (R43CA254559); the Human BioMolecular Atlas Program (HuBMAP) Integration, Visualization & Engagement (HIVE) Initiative (OT2OD026675) sub-award; the Patient-Centered Outcomes Research Institute (PCS-1609-36195) sub-award from Princess Margaret Hospital; an Elekta AB Institutional Research Grant; and the Sabin Family Foundation. Dr. Fuller was provided with direct infrastructure support by the multidisciplinary Stiefel Oropharyngeal Research Fund of the University of Texas MD Anderson Cancer Center Charles and Daneen Stiefel Center for Head and Neck Cancer, the Cancer Center Support Grant (P30CA016672), and the MD Anderson Program in Image-guided Cancer Therapy. Dr. Fuller received speaker/travel fees or support for attending meetings from Elekta AB, AAPM, the University of Alabama-Birmingham, the American Society for Radiation Oncology (ASTRO), RSNA, and the European Society for Radiation Oncology unrelated to this work. Dr. Fuller received honoraria from and provided editorial service to the American Society for Clinical Oncology unrelated to this work. Dr. Fuller received study section honoraria from the NIH unrelated to this work. Unrelated to this work, Dr. Fuller has the following patent application: U.S. Patent Application No. 16/631,662, based on International Patent Application No. PCT/US2018/042364. Dr. Fuller served as a committee member of AAPM, ASTRO, and RSNA. C.D.F. received direct industry grant support, honoraria, travel funding, and in-kind software/computer support from Elekta AB unrelated to this project. Unrelated to this work, Dr. Jhingran reports consulting fees from Genentech and patents issued for Adaptive Intracavitary Brachytherapy Applicator for Cervical Cancer MDA04-056. Barbara Marquez was supported by CPRIT and received support for attending meetings and/or travel from the University of Texas MD Anderson Cancer Center UTHealth Houston Graduate School of Biomedical Sciences Travel Award. Dr. Peterson was partially supported by NIH/NCI Cancer Center Support Grant (CCSG) P30CA016672 (Biostatistics Resource Group) and by a grant from Varian Medical Systems. Zachary Wooten was partially supported by NIH/ NCI training grant T32CA096520 and National Science Foundation (NSF) Graduate Research Fellowship Program Grant No. 1842494. Dr. Court was partially supported by CPRIT for this project and received support unrelated to this project from NCI, Varian Medical Systems, and Wellcome Trust. Not relevant to this work, L.E.C. participated in the Scientific Advisory Board for Leo Cancer Center and received stock options from Leo Cancer Center. No other disclosures were reported.